# Galaxy Formation in a Variety of Hierarchical Models

Jeremy S. Heyl[†♡], Shaun Cole[⋆♢], Carlos S. Frenk[⋆♣] and Julio F. Navarro[⋆♠]

[†]*Institute of Astronomy, Madingley Road, Cambridge CB3 0HA.*

[⋆]*Department of Physics, South Road, University of Durham, Durham DH1 3LE.*

♡ *JSHeyl@mail.ast.cam.ac.uk*

♢ *Shaun.Cole@durham.ac.uk*

♣ *C.S.Frenk@durham.ac.uk*

♠ *J.F.Navarro@durham.ac.uk*

18 August 1994

## ABSTRACT

We predict the observable properties of the galaxy population in several popular hierarchical models of galaxy formation. We employ a detailed semianalytic procedure which incorporates the formation and merging of dark matter halos, the shock heating and radiative cooling of gas, self-regulated star formation, the merging of galaxies within dark matter halos, and the spectral evolution of the stellar populations. We contrast the standard CDM cosmogony with variants of the CDM model having either a low value of $H_0$, or a low value of $\Omega$ with or without a cosmological constant. In addition, we compare galaxy formation in these CDM universes with a "cold plus hot" dark matter model (CHDM). The cosmological parameters in these models are constrained by observations of large-scale structure, light-element abundances, and globular cluster ages, while the astrophysical parameters we determine by attempting to produce the best-fit to the present-day observed $B$-band luminosity function. Having fixed the parameters in this manner we gauge the success or failure of each model by comparison with other properties of the observed galaxy population: the $K$-band luminosity function, the infrared Tully-Fisher relation, $B - K$ colours, number counts and redshift distributions. We find that although the models have some success in remedying the shortcomings of the standard CDM cosmogony, none of these new models produce broad agreement with the whole range of observations. Although the low-$\Omega$ and $\Omega + \Lambda = 1$ CDM models reduce the discrepancy between the predicted and observed Tully-Fisher relations (the main weakness of galaxy formation in standard CDM), these models predict an inverted colour-magnitude relation and do not produce an exponential cut-off at the bright end of the galaxy luminosity function. All of our models predict recent star formation in the majority of galaxies and exhibit galaxy colours bluer than observed, but this problem is far more severe in the CHDM model which produces colours about two magnitudes too blue in $B - K$. Unlike in the variants of the CDM model in the CHDM case this result is not dependent on our model of stellar feedback, but is instead directly caused by the late epoch of structure formation in this model. We discuss several potential refinements to the galaxy formation recipe: the inclusion of metallicity effects, non-local feedback, inhibited star formation in cooling flows and an initial mass function that varies in time and space.

# 1   INTRODUCTION

Studies of galaxy formation have progressed steadily over the past few years on three broad fronts: observations, numerical simulations and semianalytic treatments. On the observational front, photometric and spectroscopic data on faint galaxies and quasars suggest that the process of galaxy formation may be accessible to observation with existing techniques. For example, intense protogalactic activity seems to be occurring between redshifts 1 and 3. At this epoch, the amount of neutral hydrogen inferred in damped Lyman-$\alpha$ systems is comparable to the present mass density in stars, suggesting that these clouds might contain the raw material for most of the stars seen in galaxies today (Lanzetta *et al.* 1994). The abundance of quasars peaks in this redshift interval, signalling strong evolutionary processes (Boyle *et al.* 1990, Green 1991, Hewett, Foltz & Charree 1993) The total flux from faint blue galaxies in deep CCD counts implies intense star formation activity which may account for the production of a substantial fraction of the heavy element content of galaxies (Cowie 1989) and perhaps also for a similar fraction of their stellar content. Although the redshift range at which this flux is emitted is still undetermined, it is quite possibly near $z = 1$. Even at redshifts less than 1, there appear to be symptoms of ongoing galaxy formation, manifest in the seemingly rapid evolution in the luminosity function and the colours of galaxies, as well as in their mix of morphological types (Cowie *et al.* 1988, Lilly 1993, Ellis *et al.* 1994; Butcher & Oemler 1978, Oemler 1994, private communication).

Theoretical studies of galaxy formation have also progressed at a rapid rate, as semian-alytic modelling and numerical simulation of the relevant gravitational, gas dynamical, and radiative processes become increasingly sophisticated. Although N-body/gas dynamical simulations of large cosmological volumes do not yet have sufficient resolution to follow galaxy formation in detail, simulations of small volumes and of individual objects have shown that in hierarchical clustering theories, galaxies are assembled through mergers of dark matter halos within which gas rapidly cools and condenses, in the manner envisaged by White & Rees (White & Rees 1978, Katz, Hernquist & Weinberg 1992, Navarro & White 1993, Cen and Ostriker 1993, Katz 1992, Evrard, Summers & Davis 1994, Navarro, Frenk & White 1994a,b). The detailed mode and epoch of galaxy formation depends on the nature of the assumed primordial fluctuations. For example, in the standard cold dark matter (CDM) cosmogony , the paradigm of hierarchical clustering, galaxy formation activity peaks at relatively recent epochs, $z \lesssim 2$ (Davis *et al.* 1985, Frenk *et al.* 1988).

Semianalytic models provide a powerful tool to explore the validity of various physical assumptions and simplifications, and have been considerably extended in several recent papers (Cole 1991, White & Frenk 1991, Lacey & Silk 1991, Kauffmann, White & Guiderdoni 1993, Kauffmann, Guiderdoni & White 1994, Cole *et al.* 1994). The goal is to construct "ab initio" models in which the growth of structure arising from an assumed spectrum of primordial density perturbations is represented by a set of simple rules which encapsulate our current understanding of gravitational clustering, radiative hydrodynamics, star formation and feedback, and the ageing of stellar populations. The outcome is a set of predictions for the observable properties of the galaxy population –abundance, luminosities, colours, and circular velocities- as a function of time.

Although the detailed implementation of the rules differs somewhat in different studies, there seems to be general agreement regarding the basic features of the standard CDM model. Its successes and failures are summarized in Cole *et al.* (1994, hereafter Paper I). There it was shown that the most successful model requires star formation to be strongly regulated by feedback from supernovae and evolving stars and galaxy mergers to play a central role. With these ingredients, it is possible to construct a "fiducial model" that predicts B-band and K-band luminosity functions in general agreement with observations (although the faint-end slopes are slightly steeper than observed in the field); acceptable stellar mass-to-light ratios; a wide range of galaxy colours and a colour-luminosity relation with the correct sign; star formation rates similar to those observed; and faint number counts and associated redshift distributions in excellent agreement with observations.

The fiducial model suffers from two major shortcomings: (i) it does not produce galaxies as red as many observed ellipticals (by about 0.4 magnitudes in $B - K$) and (ii) the zero-point of the $I$-band "Tully-Fisher" relation – the correlation between the $I$-magnitude and the rotation velocity of disk galaxies – is about 2 magnitudes too faint. The first problem can be traced to the fact that standard population synthesis models require more time to generate sufficiently bright red stars than is available in the fiducial CDM model. The second problem arises because a CDM model contains an excessive number of dark galactic halos with circular velocities comparable to that of the Milky Way.

The inability to produce a fully successful model of galaxy formation may be due to an incorrect choice of cosmological parameters or to inadequacies in the modelling of the physics of galaxy formation. The aim of this paper is to explore the first of these possibilities. Thus, we retain the basic astrophysical framework of Paper I (although we reserve the freedom to adjust parameters appropriately), and apply it to a variety of alternative cosmological models. In an attempt to remedy the colour problem, we consider models with a longer timespan than CDM (by lowering the values of $H_0$ and $\Omega$) and, in an attempt to remedy the Tully-Fisher discrepancy, we consider models that produce a lower abundance of dark galactic halos (by lowering $\Omega$ or by assuming a mixture of cold and hot dark matter). Our main result is that none of these alternatives provides a satisfactory resolution to the problems affecting the fiducial model and, in many cases, they do not even share some of its successes. We are therefore led to the conclusion that some of the astrophysical processes included in our model require revision.

The remainder of this paper is organized as follows. In Section 2, we review the recipe for galaxy formation developed in Paper I and introduce the cosmological models explored in this paper. In Section 3, we present the results of our new calculations and in Section 4 we discuss their implications.

## 2  THE METHOD

### 2.1  Astrophysical Parameters: Mergers and Star Formation

The method we use to model the formation and evolution of galaxies is laid out in detail in section 2 of Paper I. Here we summarize the main features of this procedure and define the parameters that specify our model of star formation and of the merging of galaxies within a common dark matter halo.

We follow the dynamical evolution of the population of dark matter halos using the block model of Cole & Kaiser (1988; see also Cole 1991). This is an approximate Monte Carlo implementation of the analytic description of halo merging based on the extension of the Press-Schechter theory developed by Bond *et al.* (1991), Bower (1991) and Lacey & Cole (1993). The only input to the block model is the linear power spectrum normalized to the present epoch and a density threshold for the formation of a halo, $\delta_c$, calculated from the collapse of a uniform spherical overdense region. ($\delta_c$ increases with redshift and depends on both $\Omega$ and $\Lambda$.) This analytic description has recently been shown to be in good agreement with the evolution seen in large, fully non-linear, N-body simulations (Kauffmann & White 1993, Lacey & Cole 1994). The basic Press-Schechter formalism can be applied to models with $\Omega < 1$, as detailed in Lacey & Cole (1993), and extended to models with $\Lambda \neq 0$ in an entirely analogous manner. However, it is not directly applicable when a mixture of hot and cold dark matter (CHDM) is present, because relativistic neutrinos do not cluster along with the cold dark matter on scales less than their Jeans mass. Therefore, for this model we use a constant threshold of $\delta_c = 1.686$ and adopt the evolving CHDM power spectrum parameterized by Klypin *et al.* (1993). As shown by these authors, the Press-Schechter mass function determined in this manner provides a reasonable approximation to the mass function of dark halos found in N-body simulations.

Individual halos are modelled as isothermal spheres in which any diffuse gas present when the halo forms is shock heated to the virial temperature of the halo and has initially the same $\rho \propto r^{-2}$ density profile as the dark matter. This allows us to compute the fraction of gas that can cool during the halo's lifetime by computing the radius at which the cooling time, calculated assuming primordial abundances, equals the halo lifetime. The lifetime of a halo is defined as the time elapsed since the formation of a halo and its merger with a larger structure. The gas that cools is assumed to settle on a galaxy at the centre of the halo where it can then begin to form stars. In our model, this galaxy can experience more than one episode of star formation, as further episodes may be triggered by galaxy mergers.

The transformation of the cooled gas into stars is, in our model, a self-regulating process. Star formation rates are assumed to be moderated by supernovae and evolving stars, which inject thermal and kinetic energy into the gas. This energy feedback may expel gas from the galaxy, and return it to the hot diffuse phase. The efficiency of this process is assumed to depend sensitively on the depth of the potential well in which the galaxy resides. Thus, the cool gas reservoir is continuously depleted by both the transformation of gas into stars and the reheating of gas by SN. We assume that, in a halo of circular velocity, $V_c$, the star formation rate, $\dot{m}_\star(V_c, t)$, is proportional to the current mass of cool gas, $m_c(t, V_c)$,

$$\dot{m}_\star(t, V_c) = m_c(t, V_c)/\tau_\star(V_c)$$
$$= [m_c(0, V_c) - m_\star(t, V_c) - m_{\text{hot}}(t, V_c)]/\tau_\star(V_c), \tag{1}$$

where $m_{\text{hot}}(t, V_c)$ is the mass of gas reheated by the energy released from SN, that is returned to the hot phase and $m_\star(t, V_c)$ is the mass of stars formed at time $t$ after the onset of this episode of star formation. ($m_c(0, V_c)$ is the total amount of gas that can cool in the lifetime of the halo.) We further assume that the mass of gas reheated is proportional to the mass of stars formed

$$\dot{m}_{\text{hot}}(t, V_c) = \beta(V_c)\dot{m}_\star(t, V_c). \tag{2}$$

The timescale, $\tau_\star$, as well as the ratio of the mass of gas returned to the hot phase to the mass of stars formed, $\beta$, are both assumed to depend only on $V_c$. Hence

$$m_\star(t, V_c) = \frac{m_c(0, V_c)}{1 + \beta} \left[ 1 - \exp(-(1 + \beta)t/\tau_\star) \right]. \tag{3}$$

We parameterize $\tau_\star(V_c)$ and $\beta(V_c)$ as simple power laws;

$$\tau_\star(V_c) = \tau_\star^0 \left( \frac{V_c}{300\,\text{km}\,\text{s}^{-1}} \right)^{\alpha_\star} \tag{4}$$

$$\beta(V_c) = (V_c/V_{\text{hot}})^{-\alpha_{\text{hot}}}. \tag{5}$$

The four parameters $\alpha_{\text{hot}}$, $V_{\text{hot}}$, $\alpha_\star$, and $\tau_\star^0$ then specify completely our description of star formation. The simulations of Navarro & White (1993) suggest that the values of all these parameters depend only on the fraction of SN energy devolved to the ISM in the form of gas bulk motions, $f_v$ (see their Table 2). The dependences of $\tau_\star(V_c)$ and $\beta(V_c)$ on $f_v$ and $V_c$ and the corresponding values of $\alpha_{\text{hot}}$, $V_{\text{hot}}$, $\alpha_\star$ required to fit these dependences can be found in Fig. 2 and Table 1 of Paper I.

The star formation histories computed for each galaxy according to the above prescription are converted into luminosities and colours using the stellar population synthesis model of Bruzual & Charlot (1993). Here we adopt the Scalo (1986) IMF for luminous stars with masses $0.1 < M/\text{M}_\odot < 125$. The mass in non-luminous brown dwarfs with masses $M <$ $0.1\,\mathrm{M}_\odot$ is characterized by a further parameter, $\Upsilon$, defined to be the ratio of the total mass in stars to that in luminous stars.

Table 1.

| | Cosmological Parameters | | | | | Constrained Quantities | | | |
|---|---|---|---|---|---|---|---|---|---|
| Model | $\Omega$ | $\Lambda$ | $h$ | $\sigma_8$ | $\Omega_b$ | $\Gamma$ | $\sigma_8\Omega^{0.6}$ | $\Omega_b h^2$ | $t_{\mathrm{age}}/\mathrm{Gyr}$ |
| Fiducial | 1.00 | 0.00 | 0.50 | 0.67 | 0.06 | 0.5 | 0.67 | 0.015 | 13 |
| low-$H_0$ | 1.00 | 0.00 | 0.25 | 0.67 | 0.20 | 0.25 | 0.67 | 0.0125 | 26 |
| low-$\Omega$ | 0.30 | 0.00 | 0.60 | 1.0 | 0.04 | 0.3 | 0.48 | 0.0144 | 13 |
| $\Omega + \Lambda$ | 0.30 | 0.70 | 0.60 | 1.0 | 0.04 | 0.3 | 0.48 | 0.0144 | 16 |
| CHDM | 1.00 | 0.00 | 0.50 | 0.67 | 0.06 | – | 0.67 | 0.015 | 13 |

The fate of galaxies whose halos merge is determined by a merger timescale $\tau_{\mathrm{mrg}}$. If $\tau_{\mathrm{mrg}}$ is shorter than the lifetime of the newly formed common halo then we merge the two galaxies, whereas if $\tau_{\mathrm{mrg}}$ is longer than the halo lifetime the galaxies remain distinct as either a dominant galaxy and a satellite or simply as members of a cluster or group of galaxies. Galaxy mergers within hierarchically growing halos have been studied by Navarro, Frenk & White (1994a,b). They find that the probability of a merger, and hence the appropriate value of $\tau_{\mathrm{mrg}}$, depends sensitively on the angular momentum of the galaxy's orbit, but also increases with increasing galaxy mass as expected from simple considerations of dynamical friction. Hence we parameterize this merger timescale as

$$\tau_{\mathrm{mrg}} = \tau_{\mathrm{mrg}}^0 (M_{\mathrm{halo}}/M_{\mathrm{sat}})^{\alpha_{\mathrm{mrg}}}, \tag{6}$$

where $M_{\mathrm{halo}}$ is the mass of the newly formed common halo, $M_{\mathrm{sat}}$ the mass of the halo of the satellite galaxy prior to the halo merger and $\alpha_{\mathrm{mrg}} > 0$.

## 2.2 Cosmological Background

Once the astrophysical parameters have been chosen, our model of galaxy formation is fully specified by the choice of a cosmological model, as this specifies the age and density of the universe, the initial spectrum of density fluctuations, and their growth rate. For our purposes, a cosmological model is specified by six parameters: the Hubble constant, $H_0 \equiv 100h\,\mathrm{km\,s^{-1}\,Mpc^{-1}}$; the total present mass density of the universe, $\Omega$, as well as that in baryons and relativistic particles, $\Omega_b$ and $\Omega_\nu$, respectively (all these densities are expressed in units of the critical density); the cosmological constant, $\Lambda$, in units of $3H_0^2$ (so that for a flat universe $\Omega + \Lambda = 1$); and the present linear amplitude of mass fluctuations in spheres of radius $8\,h^{-1}$Mpc, $\sigma_8$.

These parameters determine the properties of the cosmological model. The *shape* of the power spectrum of linear density perturbations is determined by $\Omega$, $h$ and $\Omega_\nu$. If $\Omega_\nu = 0$ and the initial spectrum is of the Harrison-Zel'dovich form, the shape of the power spectrum in the matter-dominated era is fully specified by the shape parameter $\Gamma = \Omega h$. The *growth rate* of perturbations depends mainly on $\Omega$ and $\Lambda$, although if $\Omega_\nu > 0$ the growth of fluctuations on small scales will be retarded. In an $\Omega < 1$ universe structure ceases to grow after a redshift $z \lesssim \Omega^{-1}$. This transition is similar, but more abrupt, when $\Lambda > 0$. Consequently, models with the same value of $\sigma_8$ (and hence the same present amplitude of fluctuations) will form galactic mass halos at higher redshift for low-$\Omega$ than for $\Omega = 1$. The spatial number density of these halos is also proportional to $\Omega$ as, for the same value of $\sigma_8$, these halos will contain some fixed fraction of the total mass. The age of the universe, $t_{\text{age}}$, is proportional to $H_0^{-1}$, while for a given $H_0$ the age increases with decreasing $\Omega$ and increasing $\Lambda$. A modest increase in $\Omega_b$ can cause a large increase in the mass of stars that form because the baryon fraction controls both the total amount of baryonic material available to form stars as well as the cooling time of this material inside dark halos.

It is not feasible to present a thorough exploration of this wide parameter space. Instead, we have chosen to apply our galaxy formation framework to four new models, which we contrast with each other and with the fiducial model of Paper I. Three of the new models are variants of the standard CDM model in which $H_0$, $\Omega$ and $\Lambda$ have been varied and the fourth is the CHDM "mixed dark matter" model advocated by Davis *et al.* (1992) and Taylor & Rowan-Robinson (1992), Klypin *et al.* (1993). These four new models span the range of currently favoured cosmological models and serve to illustrate the effects of varying each of the cosmological parameters.

The parameters of these four new models are fixed by the following observational constraints:

(i) The comparison of the galaxy peculiar velocities with the density field traced by IRAS galaxies implies $\Omega^{0.6}/b_{\text{IRAS}} = 0.86 \pm 0.15$ (Kaiser *et al.* 1991, Strauss *et al.* ??), where $b_{\text{IRAS}}$ is the bias parameter relating fluctuations in the density of IRAS galaxies to fluctuations in the underlying mass distribution. The correlation function of IRAS galaxies indicates that $b_{\mathrm{IRAS}}\sigma_8 = 0.58 \pm 0.14$ (*e.g.* Moore *et al.* 1994). Assuming that the bias parameter is independent of scale these combine to yield $\sigma_8\Omega^{0.6} = 0.5 \pm 0.15$. A very similar constraint is provided by the abundance of rich clusters, which for spatially flat universes requires $\sigma_8\Omega^{0.56} = 0.57 \pm 0.05$ (White, Efstathiou & Frenk 1993).

**Table 2.** Astrophysical Parameters

| Model | $\Upsilon$ | $\tau_{\mathrm{mrg}}^0/\tau_{\mathrm{dyn}}$ | $\alpha_{\mathrm{mrg}}$ | $\tau_\star^0/\mathrm{Gyr}$ | $\alpha_\star$ | $V_{\mathrm{hot}}/\mathrm{km\,s}^{-1}$ | $\alpha_{\mathrm{hot}}$ |
|---|---|---|---|---|---|---|---|
| Fiducial | 2.7 | 0.5 | 0.25 | 2.0 | -1.5 | 140.0 | 5.5 |
| low-$H_0$ | 2.0 | 0.5 | 0.25 | 2.0 | -1.5 | 140.0 | 5.5 |
| low-$\Omega$ | 3.0 | 2.0 | 0.25 | 2.0 | -1.5 | 140.0 | 5.5 |
| $\Omega + \Lambda$ | 2.5 | 2.0 | 0.25 | 2.0 | -1.5 | 140.0 | 5.5 |
| CHDM | 1.0 | 3.0 | 0.25 | 2.0 | -1.5 | 140.0 | 5.5 |

(ii) Galaxy clustering on large scales, as measured by the APM and IRAS surveys, favour a spectrum with more large scale power than standard CDM, $\Gamma = \Omega h = 0.2$–$0.3$ (Maddox *et al.* 1990a, Efstathiou *et al.* 1990, Saunders *et al.* 1991, Feldman, Kaiser, & Peacock 1994, Fisher *et al.* 1993).

(iii) Big-Bang nucleosynthesis (BBNS) limits on primodial light element abundances require $\Omega_{\mathrm{b}}h^2 = 0.0125 \pm 0.0025$ (Walker *et al.* 1991).

(iv) Recent estimates of the age of globular clusters require $t_{\mathrm{age}} \geq 13$ Gyr (Renzini 1986; Sandage 1993).

The parameters of the fiducial model and the four new models, together with the values of these constrained quantities, are shown in Table 1. The fiducial model uses the cosmological parameters of the standard CDM and therefore fails to satisfy the constraint on $\Gamma$. With the normalization adopted here it also predicts cosmic microwave background fluctuation that are approximately 50% smaller in amplitude than those measured by COBE (Smoot *et al.* 1992). With the exception of low-$\Omega$ the normalization of all our new models is consistent with the COBE measurements.

## 3   RESULTS

As in Paper I, we choose to assess the various cosmological models described in the previ-ous section with a host of diagnostics. We proceed as follows. The "astrophysical" parameters of Table 2 are varied until an acceptable fit to the present-day $B$-band luminosity function is found for each cosmological model. Typically, this involves choosing an appropriate value of the stellar mass-to-light ratio parameter, $\Upsilon$, in order to match the knee of the $B$-band LF; the merger rate parameters, $\tau^0_{\mathrm{mrg}}$ and $\alpha_{\mathrm{mrg}}$ (which affect mainly the bright-end of the LF and are selected to suppress the formation of ultraluminous galaxies); and the parameters characterizing the star formation rates and feedback, $\tau^0_*$, $\alpha_*$, $\alpha_{\mathrm{hot}}$, and $V_{\mathrm{hot}}$ (all of which have an appreciable effect on the faint-end slope of the LF). Although we did explore departures, we choose to retain the same values as in the fiducial model for all parameters except $\Upsilon$ and $\tau^0_{\mathrm{mrg}}$. Varying the other parameters generally had little effect or resulted in an unacceptable B-band LF. The parameters of our models are given in Table 2.

Once these parameters have been specified, each galaxy formation model is fully determined. The good agreement or otherwise of each model with our additional diagnostics (the $K$-band LF, the infrared Tully-Fisher relation, the $B - K$ colours, the $B$- and $K$-number counts, and the $N(z)$ distributions) should therefore be regarded as real successes or failures of that particular cosmogony. The first three diagnostics deal with the properties of the galaxy population at $z = 0$, while the last three probe the evolutionary properties of galaxies. In some cases, and within the context of our modelling, it proved impossible to find an adequate fit to the $B$-band LF without violating one or more of the "cosmological constraints" mentioned in the previous section. When this occurs, we have explored how these constraints may be relaxed in order to improve the agreement of the model with observations. We shall comment on this in each individual case.

### 3.1 The $B$-band and $K$-band Luminosity Functions

Figure 1 presents the luminosity functions obtained for each model. The fiducial model (*i.e.* that presented in Paper I) is a reasonable fit to both the $B$- and $K$-band data. The faint-end slope seems to be slightly steeper than the Loveday *et al.* (1992) and Mobasher *et al.* (1991) data for field galaxies, but the discrepancy is not dramatic, especially noting that LFs derived from different samples (*e.g.* the CfA redshift survey or the LF in clusters: de Lapparent, Geller & Huchra 1989; Colless 1989; Driver *et al.* 1994) tend to give steeper slopes than the data used for this comparison. The faint-end slope of the fiducial model is actually much shallower than the slope of the mass function of dark halos, an effect due largely to the strong suppression of star formation in low-mass halos. This point is especially important, for producing galaxy luminosity functions as shallow as observed is a well-known problem for hierarchical clustering theories. The good agreement at the bright-end is due partly to our moderate choice of merger rates, but also to the relatively late formation of massive halos in this model. The short lifetimes of very massive halos prevent large amounts of gas from cooling to form ultraluminous galaxies at the centre of these halos. Finally, good agreement at the knee of the LF is obtained by choosing $\Upsilon = 2.7$, which indicates that a fair amount of mass should be in the form of "dark stars". The stellar mass-to-light ratios implied by fixing $\Upsilon$ in this manner and other properties of the stellar populations of our models are summarized in Table 3.

**Table 3.** Properties of the Stellar Populations: The second and third columns are the median stellar mass-to-light ratios and present star formation rates of galaxies brighter than $M_B = -19.5$. The fourth and fifth columns give the redshifts and lookback times when half the stars in each model had formed.

| Model | $(M_\star/L_\star)/(hM_\odot/L_\odot)$ | $\dot{M}_\star/M_\odot\mathrm{yr}^{-1}$ | $z_\star$ | $t_\star/\mathrm{Gyr}$ |
|---|---|---|---|---|
| Fiducial | 17 | 4.3 | 0.87 | 8.0 |
| low-$H_0$ | 34 | 7.2 | 0.71 | 14.5 |
| low-$\Omega$ | 8.9 | 4.7 | 1.17 | 8.3 |
| $\Omega + \Lambda$ | 7.7 | 7.7 | 0.86 | 8.3 |
| CHDM | 2.3 | 4.2 | 0.23 | 3.5 |

The low-$H_0$ model is also a moderately good fit to the LF data, albeit for a slightly different choice of astrophysical parameters. However, there are more stars in this model (because of the higher $\Omega_b$) and they are proportionally much older than the stars in the fiducial model (because the age of the universe has doubled). These two effects result in very high stellar mass-to-light ratios for typical galaxies: $(M_\star/L_\star) \sim 34h(M_\odot/L_\odot)$ for galaxies brighter than $M_B = -19.5$, compared to the observed $\sim 10 - 20hM_\odot/L_\odot$ in ellipticals (Lauer 1985) and $\sim 5M_\odot/L_\odot$ in the solar neighborhood (Bahcall 1984). This we regard as a serious shortcoming of the low-$H_0$ model. Reducing the value of $\Omega_b$ to less than half that prescribed by primordial nucleosynthesis can reduce the stellar mass-to-light ratios to within the observational uncertainties. However, even with this rather ad-hoc modification the model cannot account for the zero-point in the Tully-Fisher relation or for the observed colours of galaxies, as we will show in the following subsections.

The CHDM model has the opposite difficulties. The general feature of this model is that halos of galactic size form so late that they have not had time to form enough stars by $z = 0$. Matching the knee of the *B*-band luminosity function or, equivalently, the luminosity density of the universe, requires values of $\Upsilon < 1$, which are of course unacceptable. ($\Upsilon$ must be larger than unity because it is the ratio of the total mass in stars formed in a star formation burst to the mass of "visible" stars, *i.e.* excluding brown dwarfs.) Removing the feedback from star formation altogether allows more stars to form and to form earlier, but the knee in the LF nearly disappears so that it resembles a power-law rather than a Schechter function. In addition, the faint-end slope increases markedly and the luminosity density of the universe still falls short of that observed. Only by increasing $\Omega_b$ to 0.12, twice the value allowed by Big-Bang nucleosynthesis, is it possible to provide enough fuel for star formation and to match the knee of the luminosity function.

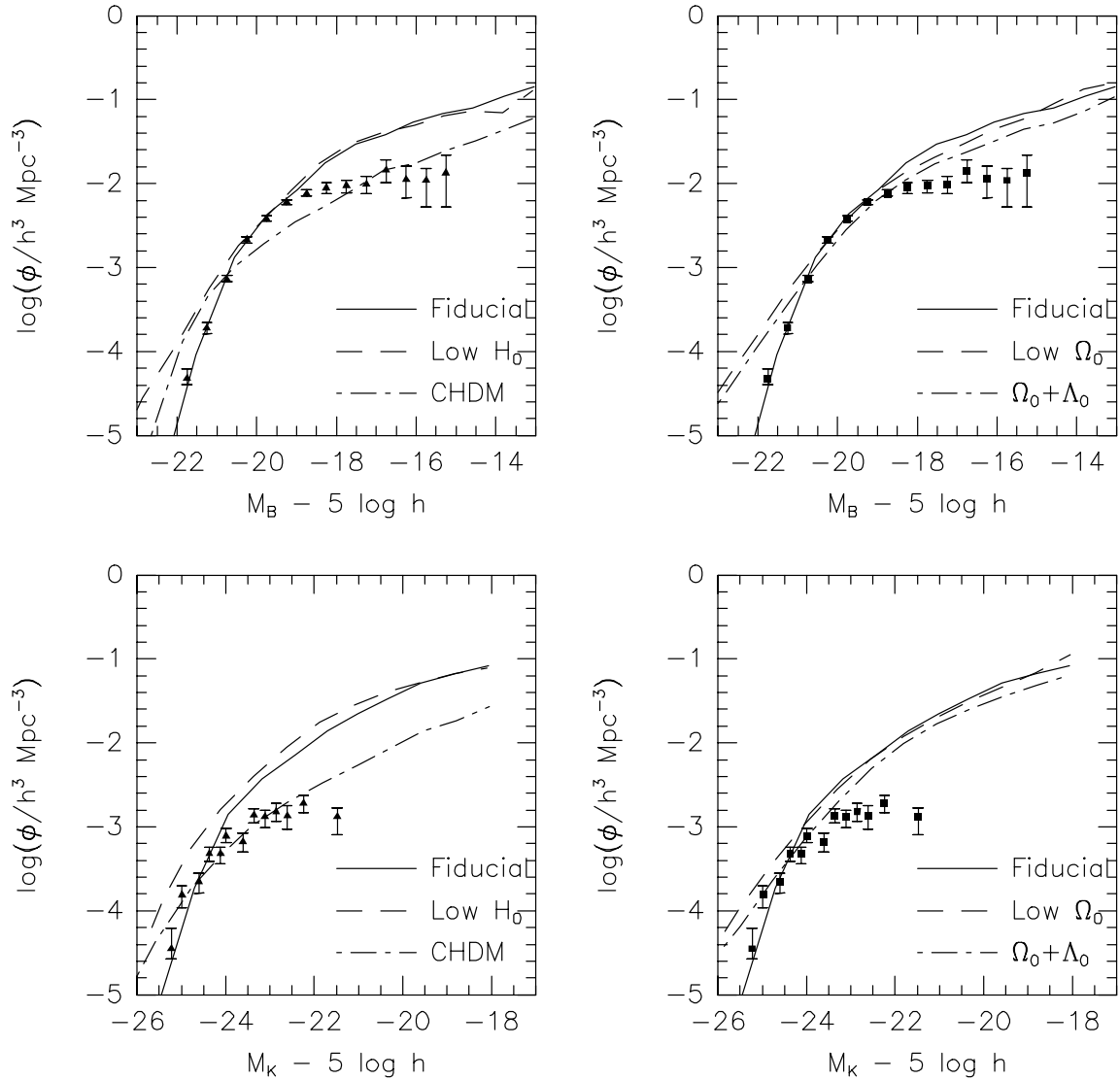

**Figure 1.** Luminosity Functions. The four panes show the luminosity functions at $z = 0$ in the B-band and K-band for the five models discussed in the paper. The points with errorbars show the Loveday *et al.* (1992) data and the Mobasher, Sharples & Ellis (1991) data. The solid line is the luminosity function of the fiducial model in all panes.

The stellar mass-to-light ratios do not pose problems to the other two cosmological models. However, both produce too many bright galaxies. Reducing the efficiency of merging has no significant effect on these galaxies, whose large luminosities are due to the fact that halos in these models have, in general, collapsed much earlier than in the fiducial model. Cooling therefore has had more time to act in massive halos, leading to the formation of overluminous galaxies. Bringing these models into agreement with observation requires postulating a star formation cutoff in very massive systems. The same problem was noticed by Kauffmann *et al.* (1993), who decided to neglect star formation in halos with circular velocities larger than about 500 km s$^{-1}$. Adopting a similar *ad-hoc* prescription here would reconcile the low-$\Omega$ and $\Omega + \Lambda$ models with the bright end of the observed LF.

## 3.2  The Tully-Fisher relation

To compare the results of our models with the observed Tully-Fisher relation we must assign rotational velocities to the "galaxies" in our model. There is no unique way of doing this because the rotational velocities of real disks are likely to be affected by the spatial distribution of the baryonic component at the centre of the dark halo (*e.g.* Persic & Salucci 1991?), an effect that is not taken into account in our model. The simplest procedure is to assign to each galaxy a rotational velocity equal to the circular velocity of the halo in which it formed. As mentioned in §1, for the fiducial model this identification results in a zero-point for the Tully-Fisher relation which is about two magnitudes fainter than observed (Figure 2). In principle, we could have adjusted the value of the stellar mass-to-light ratio parameter, $\Upsilon$, to bring the model into better agreement with the observed Tully-Fisher relation, but this would have resulted in a significant disagreement with the observed luminosity function and in a large overestimate of the luminosity density of the universe. The problem seems to be due to an overabundance of halos with circular velocities typical of bright galaxies, as noted by Lacey *et al.* (1993) and Kauffmann *et al.* (1993).

The low-$H_0$ model has the same number density of halos as the fiducial model [per $(h\mathrm{Mpc})^3$], so fitting simultaneously the galaxy luminosity function *and* the Tully-Fisher relation is not possible. The CHDM model, although better, does not resolve the problem either, despite the fact that it does not fit the present-day luminosity function, and that

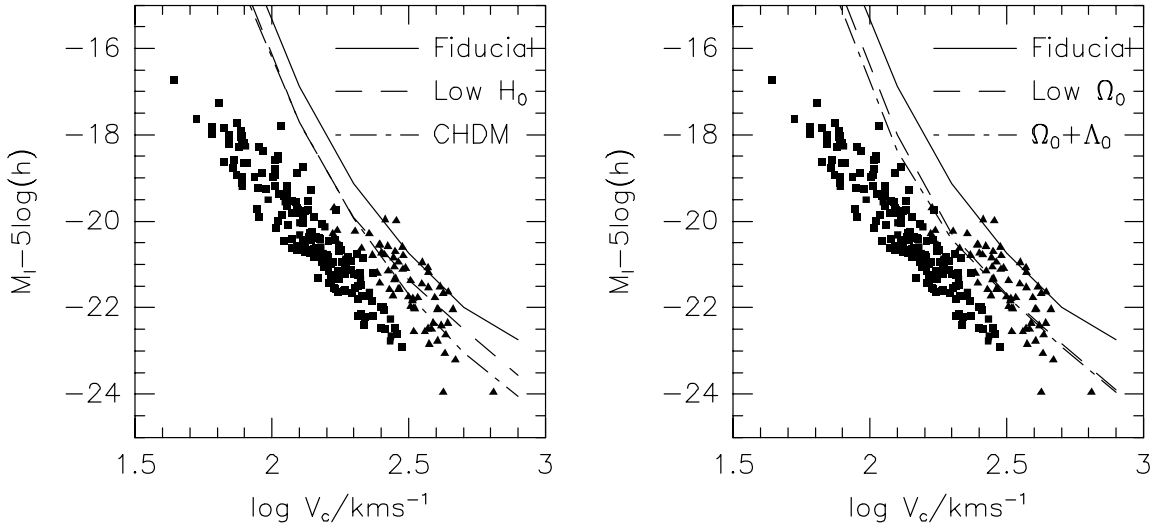

**Figure 2.** I-Band Tully-Fisher Relation. The various models are plotted as curves which trace the mean luminosity of the galaxies at a given circular velocity. The fiducial model is plotted in both panes as a solid line. The open squares correspond to a sample of spirals compiled from new and published cluster data (Young *et al.* 1994 in preparation) and the triangles to a sample of ellipticals from the Coma cluster (Lucey *et al.* 1991) which have been placed on this plane by defining an effective circular velocity in terms of the observed velocity dispersion, $V_c = \sqrt{3}\sigma_{1D}/1.1$

.

fewer galaxy-sized halos have collapsed by $z = 0$ than in the fiducial model. Both effects tend to make galaxies brighter at a given $V_c$ and to improve slightly the agreement with the observed Tully-Fisher relation. It is disappointing that increasing the value of the baryon density in order to improve the galaxy luminosity function has little effect on the Tully-Fisher relation of the CHDM model. By contrast, the number density of galaxy-sized halos is lower than the fiducial model in the cases with low-$\Omega$ and $\Omega + \Lambda$ and, therefore, their Tully-Fisher zero points agree better with observations. However, only values of $\Omega$ much lower that the one assumed in this paper would produce a zero point in full agreement with observations.

It is important to note that in all our models the discrepancy with the observed Tully-Fisher relation becomes more pronounced for low-mass halos. The predicted slope steepens below $V_{hot} = 140 \, \text{km} \, \text{s}^{-1}$, because of the strong suppression of star formation in these systems. This same effect is largely responsible for the shallow faint end slope of the luminosity function. This means that luminosity functions with faint end slopes shallower than the halo mass function *can* be obtained in our models, but only at the expense of a Tully-Fisher relation that steepens significantly towards low velocities. Such a steepening can only be compatible with the data if, because of selection effects, only the brightest galaxies have been used to define the observed Tully-Fisher relation at low $V_c$.

From this discussion it seems that, despite their wide range of parameters, none of our cosmological models can simultaneously reproduce the galaxy luminosity function and the Tully-Fisher relation. It may, however, be premature to conclude that the models are fatally flawed. The weakest link between observations and our model predictions is certainly the assumption that the rotational velocity of a galaxy is the same as the circular velocity of its surrounding halo, and there are many ways in which this identification might fail. For example, if dark halos are not well represented by singular isothermal spheres but instead possess large core radii, the rotational velocity of the galaxy's disk may not be a good indicator of its surrounding halo's $V_c$. This seems to be the case in galaxy clusters, where the velocity dispersion of the central galaxy is generally several times lower than that of the cluster itself. If disk galaxies inhabit halos that at large radii have circular velocities a factor of two larger than the disk's rotational speed, then the Tully-Fisher problem in our models would be solved. Since detailed analysis of disk rotation curves and the dynamics of satellite systems suggest that galactic halos are *not* strictly isothermal spheres, this suggestion may not be as extravagant as it appears at first sight (Persic and Salucci 1992, Ashman 1992, Zaritsky *et al.* 1993, Flores *et al.* 1993).

### 3.3 Colours

Observed broad-band colours indicate that galaxies of different magnitudes have undergone a wide variety of star formation histories. The brightest galaxies tend to be very red ($B - K > 4$), while fainter galaxies are noticeably bluer. At all magnitudes, the scatter in colours is quite large, about one magnitude in $B - K$. This is illustrated in Figure 3, where we plot data from Mobasher, Ellis & Sharples (1986) as a histogram, dividing the sample into two magnitude bins. The fiducial model fails this comparison on two counts; it does not make galaxies as red as the brightest ellipticals in this sample and it does not produce as wide a range in colours as observed. However, the trend is correct in that brighter galaxies tend to be redder than the rest. This in itself is a success for a hierarchical model in which larger systems collapse later, and comes about because stars in large galaxies today formed preferentially in smaller clumps that collapsed early and were only recently assembled into single massive objects.

It might have been expected that the low-$H_0$ model would produce a more acceptable colour distribution than the fiducial model. Indeed, galaxies are slightly redder in this model, but not nearly as red as observed. Although the universe in the low-$H_0$ model is 13 Gyr older than in the fiducial model, feedback prevents a large number of stars from forming in low-mass halos at high redshift. Star formation begins in earnest only when halos with $V_c \sim V_{\text{hot}}$ collapse and, as a result, stars form on average only about 5-6 Gyr earlier than in the fiducial model (see Table 3). The colours predicted by stellar population synthesis evolve only very slowly as stars age from $\sim 7$-8 Gyr to $\sim 12$-14 Gyr and so no major improvement in the colours results. Similarly, no significant improvements are obtained in the low-$\Omega$ and $\Omega + \Lambda$ models, where, if anything, the colour-magnitude trend is reversed. This is due to the effects of cooling and late star formation in the largest halos, as discussed in § 3.1. Not surprisingly, the CHDM model performs poorly. Although it has the same age as the fiducial model, galaxy-sized halos collapse much later and their stellar populations do not have enough time to evolve to colours comparable to those of present-day galaxies. Turning off the feedback and increasing the merger rate so as to allow stars to form in the first generation of low mass halos and then merge into more luminous systems had very little effect on the galaxy colour distribution. We also verified that this result was not sensitive to the mass resolution of the block model. Thus, the extremely blue colours of the galaxies do not depend on the details of our galaxy formation model and so are a severe problem for the CHDM model.

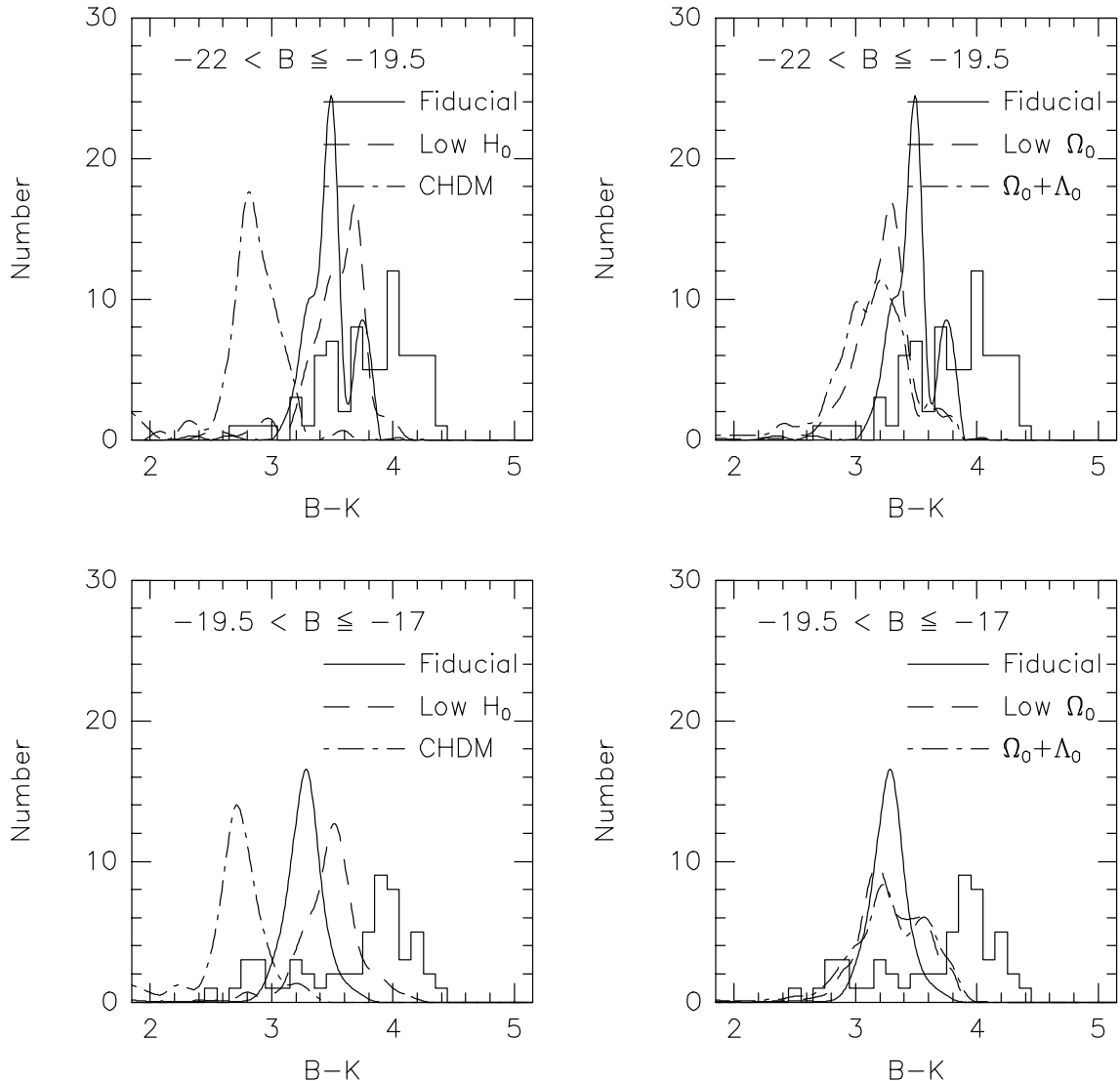

**Figure 3.** $B - K$ Colours. Each pane shows the observed colour distribution of galaxies (Mobasher *et al.* 1986) as a solid histogram and the distribution predicted by the fiducial model as a solid line. The upper panes show the distribution for bright galaxies with $-22 < M_B < -19.5$. The lower panes shows the same distribution for galaxies with $-19.5 < M_B < -17$. In all four panes the theoretical distributions have been normalised to have the same number of galaxies the observed distributions.

### 3.4   Number Counts and Redshift Distribution

Magnitude-limited number counts and redshift distributions depend on the galaxy luminosity function and its evolution and therefore probe the evolutionary properties of our models. At bright magnitudes ($B \lesssim 17$ or $K \lesssim 13$), where the $N(z)$ distribution is confined to $z \ll 1$, the number counts are determined primarily by the present day luminosity function. At fainter magnitudes, where the $N(z)$ distribution begins to pick up galaxies at higher redshifts, the counts become sensitive to several other properties of the models, such as the cosmological volume element, the galaxy colours (via K-corrections), and the genuine evolution of the galaxy luminosity function.

All our models produce too many galaxies at the brightest magnitudes ($B \lesssim 16$) as compared to the APM and EDSGC counts (Figure 4). This can be directly attributed to their present day luminosity functions which, compared to the APM luminosity function, have too many galaxies just faintwards of the characteristic luminosity and too steep a faint-end slope (Figure 1). The exception here is the CHDM model whose luminosity function also has a steep faint-end slope but does not exceed the observed luminosity function until somewhat fainter magnitudes. The counts it produces are lower than in the other models, but are still in excess of the observations at the brightest magnitudes. At fainter magnitudes, the fiducial model is roughly consistent with both the $B$ and $K$ counts and their redshift distributions. This success is due to the combination of a steep faint end slope, excessively blue colours, and significant evolution of the galaxy luminosity function.

None of the new models fare quite as well as the fiducial model despite having similar steep faint-end slopes in their $z = 0$ luminosity functions and similar galaxy colour distributions. The low-$H_0$ model predicts fewer galaxies at faint apparent magnitudes and slightly more bright ones. This feature is even more pronounced in the low-$\Omega$ and $\Omega + \Lambda$ models. In these cases the evolution of the luminosity function overcompensates for the increase in the cosmological volume element with redshift and results in fewer faint galaxies than in compared to the fiducial model. The CHDM model fails in a much more dramatic fashion. Galaxy formation occurs so late in this model that the galaxy luminosity function evolves very rapidly at low redshift. As a result, it underestimates the number of faint galaxies in both the $B$- and $K$-bands by more than a factor of four.

The redshift distributions of the faint counts provide complementary information. Figure 5 shows that the fiducial model and the low-$H_0$ model both predict $N(z)$ distributions roughly consistent with the observations. The low-$\Omega$ and the $\Omega + \Lambda$ models exhibit a small excess of high redshift galaxies in both the $B = 22$ and the $B = 24$ distributions. These tails are manifestations of the overproduction of luminous galaxies in these two models which is also apparent in their luminosity functions. By contrast, the CHDM model predicts a redshift distribution that peaks at too low redshift and has hardly any galaxies beyond $z = 0.8$, in clear disagreement with the data.

## 4   DISCUSSION

The successes and failures of the fiducial CDM model of Paper I were summarized in Section 1. We now assess, in turn, the pros and cons of each of the alternative models calculated in this paper. These models were selected specifically to find out if the deficiences of the fiducial model could be remedied within our general scheme for galaxy formation merely by changing the underlying cosmological assumptions. The parameters of these models (listed in Table 1) were chosen for consistency with recent data on galaxy clustering and peculiar velocities, Big Bang nucleosynthesis calculations, and main sequence determinations of the age of galactic globular clusters. Our strategy was to adjust the free astrophysical parameters in our scheme until the best possible agreement with the observed galaxy B-band luminosity function was obtained.

*1.)* Low-$H_0$ CDM

If $H_0$ is low, Big Bang nucleosynthesis requires a large baryon density; $\Omega_b = 0.2$ for a model with $H_0 = 25$ km s$^{-1}$ Mpc$^{-1}$. Such a large value gives rise to very efficient star formation at early times which is not significantly suppressed even when feedback effects are as strong as we have assumed. As a result, the predicted stellar mass-to-light ratios of bright galaxies turn out to be unacceptably large. This difficulty may be circumvented by violating the nucleosynthesis constraint but, if $\Omega_b$ is reduced much below 0.1, the stellar populations become too old and too faint to account for the observed abundance of bright galaxies. The best model of this kind has $\Omega_b = 0.1$, strong feedback, and a moderate amount of galaxy merging.

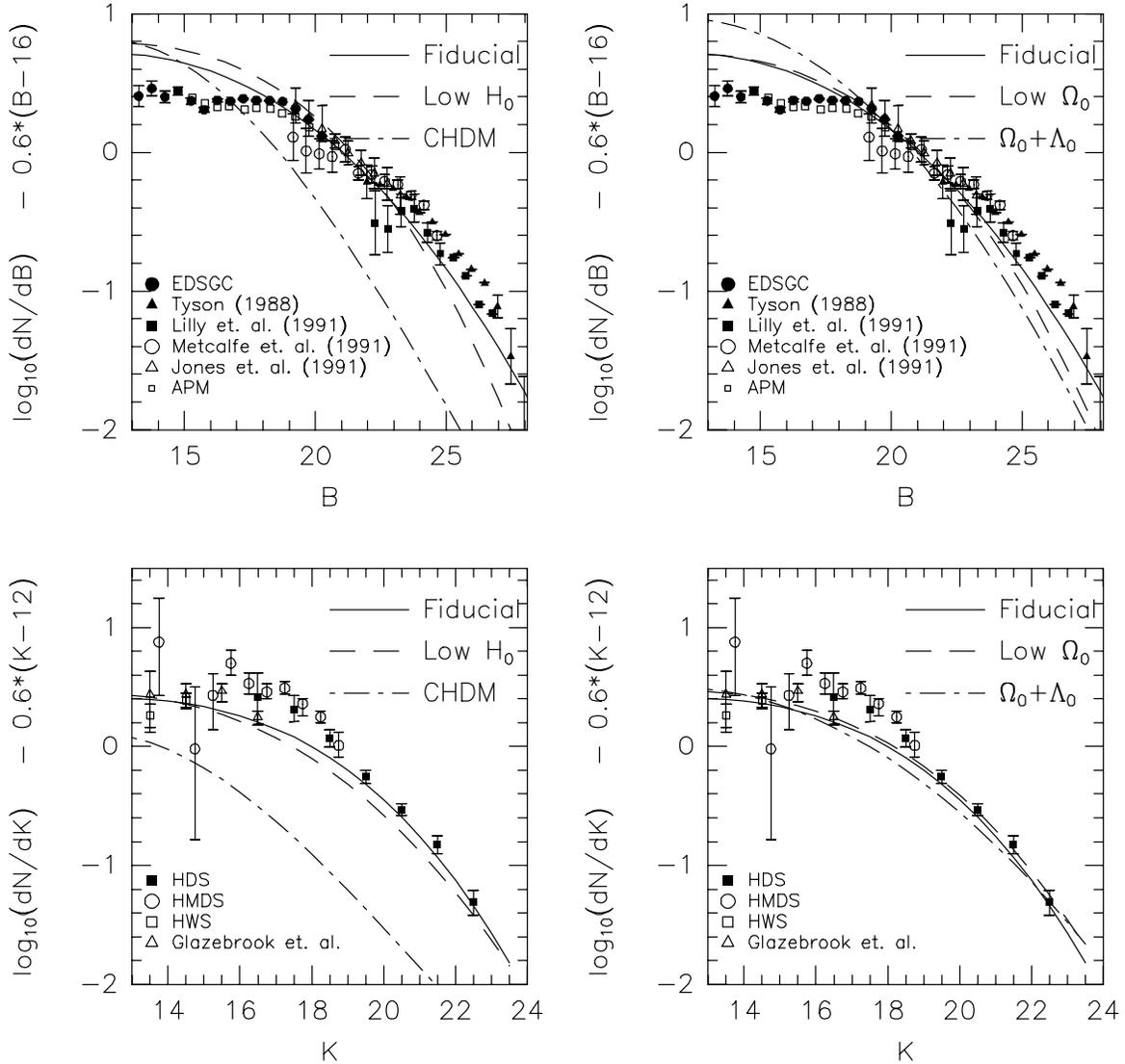

**Figure 4.** Number Counts. The upper panes show the B-band number counts for the five models discussed in the paper, and the lower panes show the K-band counts. The various polygons are the observational data, from the sources given in the key. The raw counts have been divided by a pure power law with slope 0.6, so as to expand the useful dynamic range of the figure. Thus, Euclidean number counts would appear as a horizontal line in this figure. The B-band data are taken from Maddox *et al.* (1990b), Jones *et al.* (1991), Metcalfe *et al.* (1991), Lilly *et al.* (1991), Tyson (1988) and Heydon-Dumbleton *et al.* (1989; EDSGC). Where necessary, $b_j$ magnitudes have been converted to Johnson $B$ assuming $B = b_J + 0.2$. The $K-$band data are taken from Glazebrook, Peacock & Collins (1994), the Hawaii Wide Survey (HWS), the Hawaii Medium Deep Survey (HMDS) and the Hawaii Deep Survey (HDS) as reported by Gardner, Cowie & Wainscoat (1993).

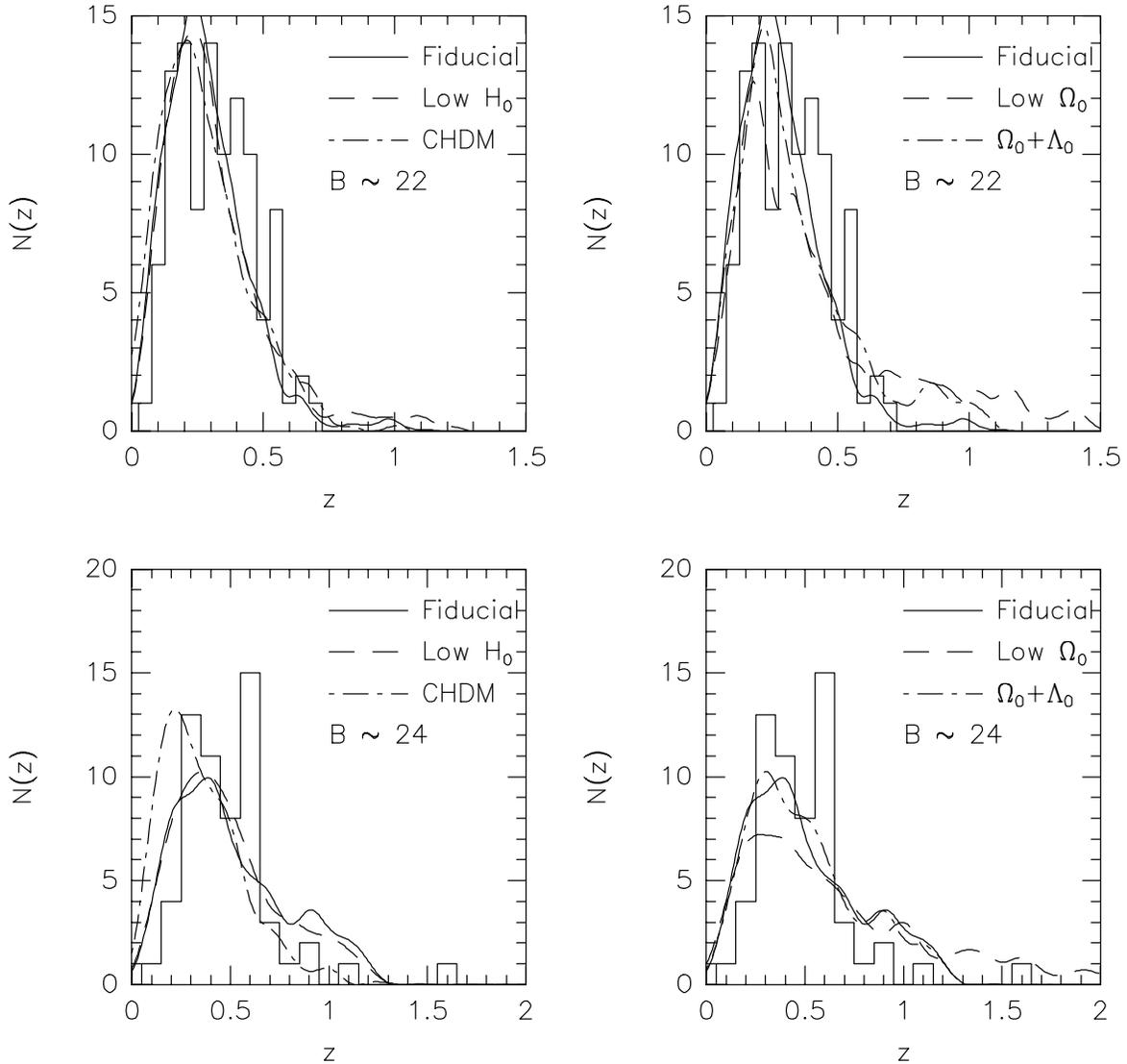

**Figure 5.** $N(z)$. The upper panes show the distribution of redshifts for a magnitude-limited sample from $B = 21$ to $B = 22.5$. For comparison, the LDSS data (Colless *et al.* 1993) are plotted as a histogram. The lower panes show the $N(z)$ distribution from $B = 22.5$ to $B = 24$ and the LDDS-2 data (Glazebrook *et al.* 1993). In all four panes the theoretical distributions have been normalised to have the same number of galaxies as the observed distributions.

The resulting luminosity function is similar to that of the fiducial model. However, the model does not fully resolve the problem which motivated it in the first place: the need to produce bright galaxies as red as many field ellipticals. Although the age of the universe is in this case 26 Gyrs, feedback effects – required to prevent an excessively large abundance of dwarf galaxies – delay the onset of star formation until relatively low redshifts and results in a paucity of very red, bright systems. Indeed, the most extreme galaxies in the model have $B - K \simeq 4$, somewhat redder than those in the fiducial model but still about 0.4 magnitudes bluer than the reddest field ellipticals.

The second main problem of the fiducial model, *i.e.* the incorrect zero point in the Tully-Fisher relation, is not resolved by lowering $H_0$. Although galaxies with a given circular velocity are about one magnitude brighter in the low $H_0$ model than in the fiducial model, they are still over 1.5 magnitudes too faint. Overall, the low $H_0$ model appears rather unattractive, especially considering the growing observational evidence in favour of a large value of $H_0$ (see *e.g.* Jacoby *et al.* 1992 and references therein).

*2.)* Low-$\Omega$ CDM

Our main motivation for examining this model was the expectation that the lower abundance of galactic halos that form in this case would be enough to bring the predicted Tully-Fisher relation into agreement with observations. This expectation was only partially fulfilled. As in the fiducial model, the predicted Tully-Fisher relation has about the observed slope for $V_c > 100 \, \mathrm{km \, s^{-1}}$, but the zero point is still about one magnitude too faint at $V_c \sim 200 \, \mathrm{km \, s^{-1}}$. Although this represents a considerable improvement over the fiducial model, it cannot be claimed as a significant success. The low-$\Omega$ model performs worse than the fiducial model on two counts: its luminosity function rolls over gently at the bright end, rather than cutting off exponentially, and the colour distribution of bright galaxies is shifted even further to the blue. The sign of the colour-magnitude relation – a notable success of the fiducial model – is inverted with brighter galaxies being bluer than fainter ones. These shortcomings can be traced to excessive cooling of gas onto large dark matter halos which form much earlier in this model than in one with a flat geometry. As noted by Kauffmann *et al.* (1994), they may be circumvented by postulating that cooling flows in large galaxies do not produce visible stars, as seems to be the case in the cooling flows inferred in the cores of rich clusters (*e.g.* Fabian *et al.* 1991). The counts of faint galaxies in the low-$\Omega$ model are almost as good as those in the fiducial model, but the excess population of bright galaxies gives rise to a significant tail of high redshift galaxies which may well be inconsistent with existing data.

*3.) $\Omega + \Lambda$ CDM*

Adding a non-zero cosmological constant to the low-$\Omega$ model has only a minor effect, although some of the small differences that there are seem to be in the right direction. The problem at the bright end of the luminosity function is slightly reduced, but the cutoff is still not as sharp as observed. The predicted Tully-Fisher relation and colour distributions change very little, but the $B$-band counts of faint galaxies drop to about a factor 2 below the data. This difference arises because the faint end slope of the luminosity function is flatter and evolves more slowly than in the fiducial model. This potential difficulty may not be too serious since, as shown in Paper I, the faint counts are rather sensitive to the assumed stellar initial mass function and to the details of the feedback prescription.

*4.) CHDM*

Like the two previous cases, a model with a mixture of cold (70%) and hot (30%) dark matter was considered in the expectation that the Tully-Fisher discrepancy of the fiducial model might be resolved. With CHDM a lower abundance of galactic halos is produced because, for a given amplitude on large scales, the power spectrum has relatively less small scale power than with CDM alone. Better agreement with the Tully-Fisher relation is indeed obtained, but the zero-point discrepancy is not fully removed. In fact, the Tully-Fisher relation in this model is virtually identical to those in the low-$\Omega$ and $\Omega + \Lambda$ models.

The reduced spectral power on galactic scales has an undesirable side effect which makes the CHDM model rather unattractive; bright galaxies form much too late to be consistent with observations. With $H_0 = 60$ km s$^{-1}$ Mpc$^{-1}$, the baryon density required by Big Bang nucleosynthesis constraints is too low to form enough bright galaxies to match the knee of the luminosity function. Even if we disregard the BBNS constraints and arbitrarily set $\Omega_b = 0.1$, the resulting luminosity function does not show the characteristic break at high luminosities. Perhaps more damning are the extremely blue galaxy colours predicted at the present epoch which, in the mean, are about 2 mag bluer than observed. The reddest objects in the model have $B - K \simeq 3.5$, one magnitude short of the reddest observed ellipticals.

These difficulties are also manifest in the counts of faint galaxies, which are a factor of 10 lower in the $K$-band than observed, and in their redshift distribution which is strongly biased towards low redshift, in strong disagreement with observations. The problem of late galaxy formation in the CHDM model thus seems unsurmountable. This conclusion is virtually independent of the details of our galaxy formation model. Even if we switch off the feedback altogether and adopt a very short star formation timescale (which produces a completely unacceptable luminosity function) bright galaxies are still much too blue.

Our failure to find a fully consistent picture of galaxy formation within currently popular cosmologies, suggests that we should look carefully at the astrophysical inputs that go into our modelling procedure. The colour problem, common to all the cases we have examined (including the long-lived low-$H_0$ model), is particularly puzzling. For carefully chosen star formation rates, the stellar population synthesis model which we use produces acceptable fits to the integrated spectral energy distributions of present day galaxies of all spectral types. However, the more realistic star formation laws in our models invariably produce intermediate age stellar populations in bright galaxies from the late infall of gas expelled from halos in the lowest level of the clustering hierarchy. (Regardless of the detailed prescription for feedback, gas must be prevented from forming stars profusely in these low-mass halos; otherwise virtually all the baryons would be turned into stars well before the present, and an unacceptably large abundance of dwarf galaxies would result.) It is possible that a star formation rate more strongly biased towards high redshift than in our models generically predict might circumvent these problems. Another possibility is that current stellar population synthesis models are predicting colours which are too blue at the 0.3 mag level in $B-K$. Such inaccuracies might arise from the treatment of the poorly understood late stages of stellar evolution (particularly the asymptotic and post asymptotic giant branch) or from the neglect of chemical evolution.

We have argued that some form of feedback is an essential requirement in any hierarchical clustering theory of galaxy formation. The ejection of gas (and metals) observed in bright ellitpicals, sometimes in the form of highly energetic superwinds (David, Forman & Jones 1991; Heckman, Armus & Miley 1990), provides an example of the sort of process which may be required. Nevertheless, there is no direct observational guidance for assuming any particular form of feedback in the highly specific conditions prevailing at high redshift. The feedback mechanism implemented in our scheme and in most other related ones is a local process where star formation is regulated *in situ*. Non-local processes such as photoionisation (Efstathiou 1992) could be important and it is not inconceivable that they could either depend on the large-scale environment or act selectively, allowing early formation in some halos and delaying it or suppressing it altogether in others. Processes of this sort might alleviate the colour discrepancy discussed above and could even give rise to "naked halos"; *i.e.* dark matter objects in which no visible galaxy ever forms. The Tully-Fisher discrepancy in the standard CDM model, and probably also in the alternative models which we have considered, could be resolved if a substantial fraction of dark halos do not harbour bright galaxies.

A further source of uncertainty in our scheme for galaxy formation in general, and in the stellar population synthesis models in particular, is the stellar initial mass function (IMF). The universality of the locally determined IMF has been a longstanding matter of much debate. Perhaps the strongest argument for a non-universal IMF comes from studies of the metallicity of the intracluster gas which seems to require a bimodal IMF in the metal-producing galaxies (Arnaud *et al.* 1992). It is not difficult to speculate on the many outcomes possible with a variable IMF. For example, an IMF biased towards massive stars in low-mass systems might alleviate the problem related to the excessive number of low-luminosity galaxies if their stellar populations have faded by the present day.

It should be clear from the above discussion that, subject to the observational constraints of large scale structure, Big-Bang nucleosynthesis, and globular cluster ages, our semianalytic recipe for galaxy formation fails to produce a fully acceptable model. Our results are quite consistent with those of Kauffmann *et al.* (1993, 1994). This agreement strengthens our conclusions since the two approaches, although similar in spirit, differ significantly in many astrophysical details. It is difficult to see how, without revision of our scheme or dramatic changes in the interpretation of observations, hierarchical models of the kind described in this paper can successfully account for the observed properties of the galaxy population. This is illustrative of the potential of the semianalytic methods we have used in this paper; they enable us to test a wide variety of models and assumptions as well as to isolate the root causes of disagreement between observations and specific cosmogonies. This, in itself, should be regarded as a success of our modelling technique, as it highlights the obstacles to be dealt with by future attempts at unravelling the process of galaxy formation in a hierarchical universe.

## ACKNOWLEDGMENTS

The authors wish to thank Alfonso Aragón-Salamanca and Steve Zepf for their considerable contribution to the construction of these galaxy formation models. JSH would like to acknowledge the Marshall Aid Commemoration Commission for support during the preparation of this work. SMC and JFN are supported by PDRA's sponsored by the PPARC.